\documentclass[epsfig,graphics,twocolumn,showpacs,floatfix,mathbbm,prl,superscriptaddress]{revtex4}
\usepackage{graphicx}
\usepackage{subfigure}
\usepackage{textcomp}
\usepackage{amssymb}
\usepackage{amsmath}
\usepackage[space]{grffile}

\begin{document}

\title[a]{Ultranarrow-Band Photon Pair Source Compatible with Solid State Quantum Memories and Telecommunication Networks}

\pacs{03.67.Hk,42.50.Ar,42.50.Dv,42.50.Ex}

\author{Julia Fekete}
\affiliation{ICFO-The Institute of Photonic Sciences, Mediterranean Technology Park, 08860 Castelldefels, Spain}

\author{Daniel Riel\" ander}
\affiliation{ICFO-The Institute of Photonic Sciences, Mediterranean Technology Park, 08860 Castelldefels, Spain}

\author{Matteo Cristiani}
\affiliation{ICFO-The Institute of Photonic Sciences, Mediterranean Technology Park, 08860 Castelldefels, Spain}

\author{Hugues de Riedmatten}
\email{hugues.deriedmatten@icfo.es}
\affiliation{ICFO-The Institute of Photonic Sciences, Mediterranean Technology Park, 08860 Castelldefels, Spain}
\affiliation{ICREA-Instituci\' o Catalana de Recerca i Estudis Avan\c cats, 08015 Barcelona, Spain}


\begin{abstract}
We report on a source of ultranarrow-band photon pairs generated
by widely nondegenerate cavity-enhanced spontaneous
down-conversion. The source is designed to be compatible with $\mathrm{Pr}^{3+}$ solid state quantum memories and telecommunication optical fibers, with signal and idler photons close to $606\, \mathrm{nm}$  and $1436\, \mathrm{nm}$, respectively. Both photons have a spectral bandwidth around $2\, \mathrm{MHz}$, matching the bandwidth of $\mathrm{Pr}^{3+}$ doped quantum memories. This source is ideally suited for long distance quantum communication architectures involving solid state quantum memories.
\end{abstract}

\maketitle 
The interaction between quantum light and matter has
important implications in quantum information science because it
forms the basis of quantum memories (QM) for light \cite{Hammerer2010, Simon2010,Lvovsky2009,Sangouard2011}. For
applications in long distance quantum communication networks, it
is required that QMs are connected to optical fibers. However,
most QMs absorb photons at wavelengths very far from
telecommunication wavelengths. To overcome this limitation,
possible solutions include quantum frequency conversion
\cite{Radnaev2010} or entanglement between QMs and
telecommunication photons via nondegenerate photon pair sources
as proposed in \cite{Simon2007a}.

Rare-earth doped solids have shown very promising characteristics
as solid state QMs \cite{Tittel2010,Sangouard2011}. Short lived
entanglement between atomic excitations stored in rare-earth
crystals and photons at telecommunication wavelength has been
demonstrated recently
\cite{Clausen2011,Saglamyurek2011}. 
Among the many rare-earth elements, praseodymium
($\mathrm{Pr}^{3+}$) doped solids possess, so far, the best
demonstrated properties in terms of quantum storage. For example,
a very high storage efficiency (up to $69\%$) has been
demonstrated \cite{Hedges2010}. In addition, in contrast to the
materials used in \cite{Clausen2011,Saglamyurek2011},
$\mathrm{Pr}^{3+}$  has a ground state structure with 3 long lived
hyperfine levels, enabling long term light storage in spin states,
with storage times $>$ 1 s
demonstrated for coherent light pulses \cite{Longdell2005}. 

Despite these exceptional properties, there is, so far, no quantum
light source compatible with $\mathrm{Pr}^{3+}$ QMs. Instead, all
experiments were done using either strong
 \cite{Longdell2005,Afzelius2010,Heinze2011} or weak
\cite{Hedges2010,Sabooni2010, Gundogan2012} coherent states of
light.  The main challenge in realizing such a light source is that
the bandwidth of the memory is limited to a few MHz by the excited
state spacing of the ions. In addition, as mentioned above, it
would be desirable to have a photon pair source with one photon
compatible with the QM and the other photon at telecommunication
wavelength to minimize losses for long distance fiber transmission
\cite{Simon2007a}.

Spontaneous parametric down-conversion (SPDC) is an interesting
candidate for realizing such a quantum light source. It is a standard
technology and the phase-matching can be tailored to allow for the
creation of photons at very different wavelengths. However, the
frequency spectrum obtained in down-conversion (typically
$>100\,\mathrm{GHz}$) is several orders of magnitude larger than
the bandwidth required to interact with $\mathrm{Pr}^{3+}$ doped
QMs. One possibility for overcoming this problem is to spectrally
filter the SPDC output \cite{Haase2009}. This solution requires
SPDC sources with large spectral brightness \cite{Haase2009}, such
as nonlinear waveguides \cite{Clausen2011,Usmani2012} and several
stages of filtering. 
Another possibility is to place the crystal inside an optical
cavity, which enhances the probability of generating a photon pair
in the resonant mode with respect to the single-pass case.
Cavity-enhanced SPDC was first demonstrated in 1999 \cite{Ou1999}
and since then many experiments have been performed in various
cavity designs
\cite{Wang2004,Kuklewicz2006,Pomarico2009,Foertsch2012,Chuu2012}.
Recently, atom-resonant photon pairs have been generated
\cite{Bao2008,Scholz2009,Wolfgramm2011} and used for quantum storage \cite{Zhang2011} and
quantum metrology application \cite{Wolfgramm2013}. In most
cavity-enhanced down-conversion experiments so far, the photons
were frequency degenerate or close to degeneracy (with the
exception of \cite{Foertsch2012} where the wavelength difference
was $100\, \mathrm{nm}$). In nondegenerate cavity sources the
enhancement of photon pair generation probability occurs only for
a limited set of longitudinal modes, called clusters
\cite{Jeronimo-Moreno2010,Pomarico2009}. This clustering effect
has been recently proposed for engineering single-mode narrow-band
sources \cite{Pomarico2012,Chuu2012}.

In this Letter, we report on an ultranarrow-band widely nondegenerate photon pair source compatible with $\mathrm{Pr}^{3+}$ doped QMs and with telecom optical fibers. The photon pairs are obtained by cavity-enhanced SPDC and have a wavelength near $606\, \mathrm{nm}$ (the resonant wavelength of  $\mathrm{Pr}^{3+}:\mathrm{Y}_2\mathrm{SiO}_5$ crystals, signal photon) and $1436\, \mathrm{nm}$ (idler photon). The measured correlation time of the pairs is $104\, \mathrm{ns}$, the longest reported for a SPDC source to our knowledge, leading to single photon spectral bandwidths of $1.6-2.9\, \mathrm{MHz}$, matching the bandwidth of $\mathrm{Pr}^{3+}$ doped QMs. In addition, we show a strong reduction of the number of longitudinal modes emitted by the cavity due to the clustering effect,
which facilitates the filtering to obtain single-mode operation. 
Indeed, by  inserting a properly designed filter for the
signal photons, we obtain strong evidence of single-mode operation.

\begin{figure}[!hbt]
    \centering
    \includegraphics[width=8cm]{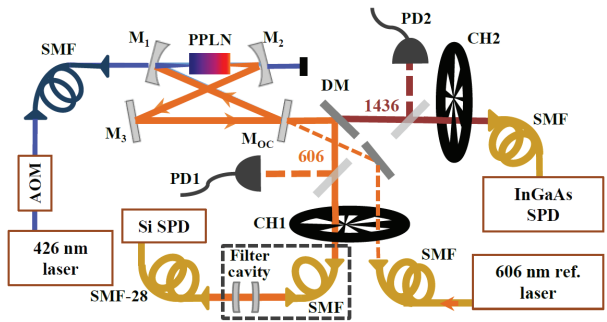}
    \caption{Experimental setup. AOM, acousto-optic modulator; SMF, single-mode fiber; M$_{1,2,3}$ highly reflective mirrors; M$_\mathrm{OC}$ output coupler mirror; DM, dichroic mirror;
    PD1,2, photodiodes; CH1,2, choppers; SPD, single-photon detectors.\label{fig:setup}}
\end{figure}

The scheme of the photon pair source is shown in
Fig.~\ref{fig:setup}. The pump for the down-conversion is a $426.2\,
\mathrm{nm}$ single-frequency continuous-wave laser system
(Toptica TA SHG). The pump light is passing through an
acousto-optic modulator (AOM) that allows for varying the pump
power incident on the cavity. After the AOM the light is coupled
into a single-mode fiber (SMF) for spatial mode cleaning. The SPDC
process is based on type I quasi-phase-matching in a $2\,
\mathrm{cm}$ long periodically poled lithium niobate (PPLN) crystal ($5\,
\%$ $\mathrm{MgO}$ doped PPLN with $16.5\,\mathrm{\mu m}$ poling
period, AR coated for $426$, $606$, $1436\, \mathrm{nm}$, provided
by HCPhotonics Corp.) The bow-tie cavity surrounding the crystal
has finesse $\mathcal{F}\approx200$, and free spectral range
$FSR\approx 414\, \mathrm{MHz}$. The mirror reflectivities  for
signal and idler wavelength are $99.99\%$ for M$_{1,2,3}$, and
$98.5\%$ for the output coupler (M$_\mathrm{OC}$, Layertec
GmbH).

In order to obtain enhancement of the photon pair generation
probability, resonance for both the signal and idler frequencies has to be ensured. In
addition, the signal frequency has to be resonant with
$\mathrm{Pr}^{3+}:\mathrm{Y}_2\mathrm{SiO}_5$ crystals. To achieve these conditions a complex
lock system was implemented. A reference beam at $606\,
\mathrm{nm}$, resonant with $\mathrm{Pr}^{3+}$ ions
\cite{Gundogan2012}, is coupled into the cavity through
M$_\mathrm{OC}$, copropagating with the pump beam. The reflection
from the cavity is used for locking the cavity length at resonance
with the signal frequency, by means of a piezo element moving
M$_\mathrm{OC}$. A reference beam at the idler frequency is
obtained by difference-frequency generation between the pump
($426\,\mathrm{nm}$) and the $606\, \mathrm{nm}$ reference beam
and is used  for locking the pump frequency to maintain double
resonance. The lock system is acting during $45\%$ of the total
operation time, alternating with the measurement time, separated
by mechanical choppers  (CH1,2). During the measurement time, no
classical reference signal beam is present in the cavity to
protect the single-photon detectors (SPD) and avoid excessive
noise. The down-converted photons are spectrally filtered, fiber
coupled (SMF-28: single-mode at $1436\, \mathrm{nm}$, and slightly
multimode at $606\, \mathrm{nm}$) and detected by SPDs (Si SPD:
Perkin Elmer SPCM-AQR-16-FC with detection efficiency,
$\eta_{\mathrm{det,}606}\approx 60\,\%$; InGaAs SPD: IdQuantique
id220 used in free-running mode with
$\eta_{\mathrm{det,}1436}\approx 10-20\,\%$). The measured
transmission from the output of the cavity to the SPDs is 0.46
(0.31) for the signal (idler) photons.

\begin{figure}[!hbt]
    \centering
    \includegraphics[angle=-90,width=8.0cm]{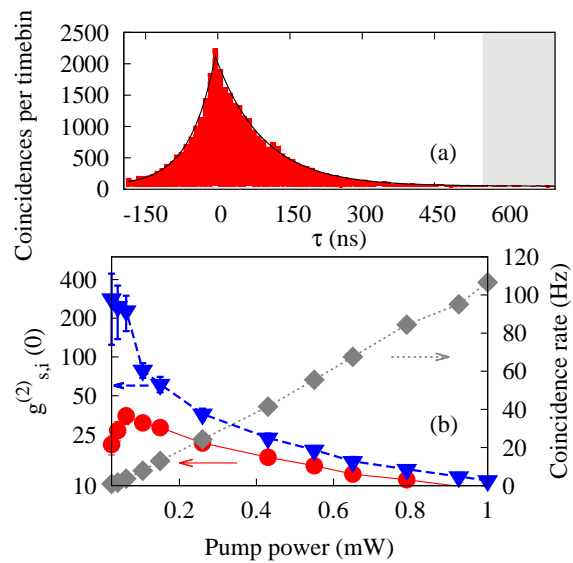}
    \caption{Measured correlation functions of the unfiltered photon pair source. (a) $G^{(2)}_{s,i} (\tau)$ function (measured at $130\,\mathrm{\mu W}$ pump power, $\eta_{\mathrm{det,}1436}=10\,\%$, $43\, \mathrm{min}$ integration time, $10\, \mathrm{ns}$ time-bin size). The dark count rate (light-gray bars of values below 50) is obtained by blocking the InGaAs SPD.
$g^{(2)}_{s,i} (0)$ is calculated using the fit function (solid line) as the peak value divided by the accidental region (shaded area) average count rate level. (b) Power dependence of  $g^{(2)}_{s,i} (0)$ without and with dark count subtraction (left axis, circles and triangles, respectively) and coincidence count rates (right axis, diamonds).
      \label{fig:MM-G2}}
\end{figure}

The quantum state of the emitted photon pairs can be analyzed by
the second-order cross-correlation function between the signal and
idler fields, $G^{(2)}_{s,i} (\tau)$ - see appendix \ref{appdx} for details. To do so, the frequency nondegenerate
photons of the pair are spatially separated by a dichroic mirror
(DM) and their arrival times at the detectors are recorded by a
time-to-digital-converter card (Signadyne). To create the
correlation function, the arrival-time differences ($\tau$) are
plotted in a histogram, as shown in Fig.~\ref{fig:MM-G2}(a).

We observe a FWHM correlation time of  $104\, \mathrm{ns}$, which,
to our knowledge, is the highest value reported from a SPDC source
so far. The double exponential decay of $G^{(2)}_{s,i} (\tau)$ is
due to the photon lifetime inside the cavity and allows for the
estimation of the cavity linewidth, $\Delta \nu$. Fitting
$\exp(-2\pi \Delta \nu \tau)$ on the two sides of the correlation
function results in $\Delta \nu = 1.7\, \mathrm{MHz}$ at $1436\,
\mathrm{nm}$ and $2.9\, \mathrm{MHz}$ at $606\, \mathrm{nm}$. The
asymmetry is a clear sign of the difference of the finesse for the
two wavelengths which can be explained by a slight difference in
the reflectivities of the mirrors or the crystal surfaces. From these values we infer internal
cavity losses of $1-2.5\,\%$ and cavity escape efficiencies of
$\sim0.6-0.4$ \cite{Wolfgramm2011a}.

To further characterize the source, the histograms were measured
at various pump power levels. The number of total coincidences
within a $500\,\mathrm{ns}$ time window increases linearly with
pump power [see Fig.~\ref{fig:MM-G2}(b)]. The detected pair
production rate after dark count subtraction is $C=100\,
\mathrm{Hz/mW}$. For applications in quantum information science,
another critical parameter is the normalized cross-correlation
function at zero delay, $g^{(2)}_{s,i} (0)$, which is also plotted
in Fig.~\ref{fig:MM-G2}(b)  as a function of pump power. In order
to obtain good fidelities in quantum protocols, high
$g^{(2)}_{s,i} (0)$ is needed. As expected, $g^{(2)}_{s,i} (0)$ is
inversely proportional to the pump power \cite{Foertsch2012}.
Without subtracting any background, we reach values of
$g^{(2)}_{s,i} (0)$ up to 35, significantly higher than the
classical threshold of 2 for two-mode squeezed states. For a pump
power of $1\,\mathrm{mW}$, we still observe $g^{(2)}_{s,i}
(0)=9.3$. At very low power (below $100\,\mathrm{\mu W}$) the
measurements are limited by dark counts, reducing $g^{(2)}_{s,i}
(0)$ \cite{Sekatski2012}. The $g^{(2)}_{s,i} (0)$ with dark count
subtraction (also plotted) approaches a value of 284 at
$24\,\mathrm{\mu W}$, showing the high purity of our photon pair
source.

As a possible application of the photon pair source, the telecom
photons could be used to herald the presence of the signal photons to be absorbed by the QM. To give an estimate
for the heralding efficiency ($\eta^H$), the singles count rate at
the InGaAs SPD ($S_{1436}$) 
has to be taken into account,
resulting in $\eta^H=C/S_{1436}/\eta_{\mathrm{det,}606}= 13\,\%$.

\begin{figure}[!hbt]
    \centering
    \includegraphics[angle=0,width=8cm]{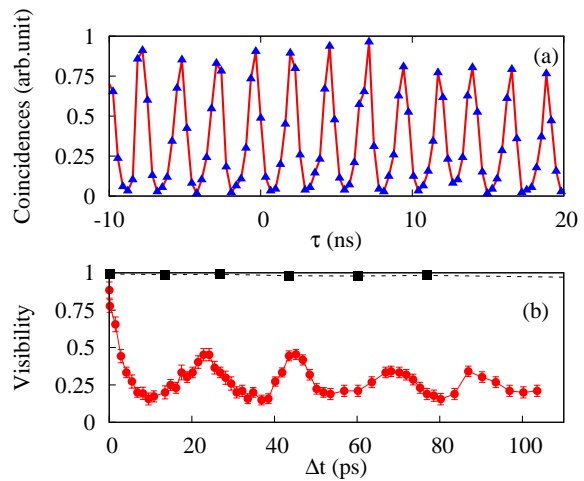}
    \caption{Temporal oscillations in the correlation functions for the unfiltered photon pair source. (a) Zoom of the $G^{(2)}_{s,i} (\tau)$ function of Fig.~\ref{fig:MM-G2}(a) to the $-10-20\, \mathrm{ns}$ region evaluated with $325\, \mathrm{ps}$ time-bin size. (b) $|g^{(1)}_{s} (\Delta t)|$ function (circles). As reference, the visibility for classical light is shown (squares).
\label{fig:MM-osc}}
\end{figure}

We now investigate the spectral content of the emitted photons.
Compared to the single-pass SPDC, where the spectrum is determined
by the phase-matching bandwidth ($\mathrm{FWHM}\approx 80\,
\mathrm{GHz}$ in our case), for cavity-enhanced SPDC the spectrum
is modified by the cavity modes. Since the signal and idler have
different dispersion characteristics, the respective $FSR$s also
differ, and the joint spectral amplitude will consist of spectral
modes restricted to certain clusters
\cite{Jeronimo-Moreno2010,Pomarico2012}. The cluster spacing and
width are determined by the dispersion relations and the cavity
geometry. A first insight on the spectral content of the emitted
photons can be inferred from the $G^{(2)}_{s,i} (\tau)$ function.
As shown in Fig.~\ref{fig:MM-osc}(a), the histogram taken with
$325\, \mathrm{ps}$ timebin size shows an oscillatory behavior.
This is the consequence of the presence of multiple spectral modes
\cite{Wang2004}. The periodicity follows the cavity round-trip
time ($T = 1/FSR = 2.4\, \mathrm{ns}$), as the photon wave packet
incorporates the beating from several frequency components spaced
by the $FSR$. From the temporal width of the peaks ($\sim
880\,\mathrm{ps}$) and taking into account the time resolution of
the detection system ($\sim 685\,\mathrm{ps}$), we infer that the
spectrum is composed of clusters containing around 4 longitudinal
modes (see appendix \ref{appdx}). This technique is,
however, limited to a few GHz by the detection time resolution.

 To further investigate the
presence of multiple modes we measured the first order
autocorrelation (coherence) function of the signal photons,
$V=|g^{(1)}_{s} (\Delta t)|,$ in a $100\, \mathrm{ps}$ range using
a Michelson interferometer \cite{Wang2010}. This was obtained by
measuring the visibility $V$ of the interference fringes as a
function of the time difference $\Delta t$ between the two arms of
the interferometer. To vary the path length difference, one of the
mirrors was placed on a translation stage together with a piezo
element for fine tuning. The results are shown in
Fig.~\ref{fig:MM-osc}(b).

\begin{figure}[!hbt]
    \centering
    \includegraphics[angle=-90,width=8cm]{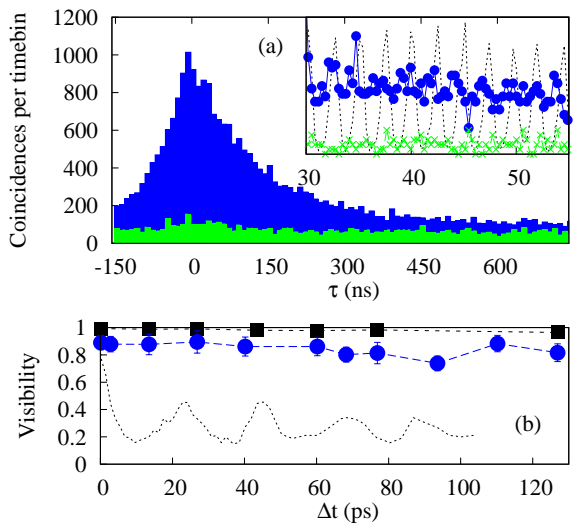}
    \caption{Measured correlation functions of the cavity output when the signal was spectrally filtered. (a)
$G^{(2)}_{s,i} (\tau)$ function ($1.7\, \mathrm{mW}$ pump power,
$\eta_{\mathrm{det,}1436}=20\,\%$, $60\, \mathrm{min}$ integration
time,  $10\, \mathrm{ns}$ time-bin size). The noise level [light-gray (green in color version) bars] was measured with FC off-resonant with the signal. The inset
is a zoom to the $30-55\, \mathrm{ns}$ region evaluated with
$325\, \mathrm{ps}$ time-bin size. The dashed line corresponds to
the unfiltered case for comparison. (b) $|g^{(1)}_{s} (\Delta t)|$
function measured with $606\, \mathrm{nm}$ down-converted photons
after dark count subtraction (circles). As reference, the
visibility for classical light (squares), and for the unfiltered
cavity (dotted line) are plotted.
     \label{fig:SM}}
\end{figure}

The oscillations that we observe in the $|g^{(1)}_{s} (\Delta t)|$
function can be understood from the clustering effect. From
the $22.5\, \mathrm{ps}$ oscillation period, we can infer the
presence of clusters separated by $44.5\, \mathrm{GHz}$. From the
calculated phase-matching bandwidth, we infer that at most 3
clusters are present. During the locking period the main cluster is centered in the
vicinity of the peak of the phase-matching curve. The contrast of
the oscillations ($\sim$ 0.4) is compatible with a suppression of
more than 80 $\%$ of the side clusters. The interference
visibility for the oscillation peaks is reduced to $\sim45\,\%$, which can be explained by the presence
of broadband noise that was not eliminated properly.

The multimode cavity could in principle be used as the source for
storage of several longitudinal modes within the inhomogeneous
absorption line of a $\mathrm{Pr}^{3+}$ QM, to implement frequency
multiplexing. Nevertheless, in absence of a frequency multimode
QM, it is important to show that our source can also operate in
the single-mode regime. To achieve this, we inserted a filter
cavity (FC) in the $606\,\mathrm{nm}$ arm, designed to suppress
all modes around the fundamental one. The $FSR$ was chosen such
that the transmission peaks do not coincide with the neighboring
clusters ($\Delta\nu=80\, \mathrm{MHz}$, $FSR=16.8\,
\mathrm{GHz}$). The FC transmission for long term measurements was
$11\,\%$, including a peak transmission of $22\,\%$
(fiber-to-fiber) and a further drop of $50\,\%$ due to frequency
drifts of the only passively stabilized FC. For this measurement,
a single-mode fiber was inserted in the signal arm
[with reduced fiber coupling compared to the unfiltered cavity
experiments ($\eta_C=0.5$ and 0.8, respectively)].

The measured $G^{(2)}_{s,i} (\tau)$ function of the signal
transmitted by the FC and the unfiltered idler is plotted in
Fig.~\ref{fig:SM}(a). We obtain a raw $g^{(2)}_{s,i} (0)$ of 9.
Note that a significant part of accidental coincidences are due to
the fact that only the signal photon is filtered. The number of
coincidences ($500\, \mathrm{ns}$ window) with the background
level subtracted is $2.9\, \mathrm{Hz/mW}$. Comparing this with the
unfiltered cavity--correcting for the different experimental
conditions--we find a reduction of $\sim4.5$ for the coincidence rate
which is consistent with the number of modes estimated from the
unfiltered experiment.

The suppression of the oscillations at $2.4\, \mathrm{ns}$ for the
filtered signal [inset of Fig.~\ref{fig:SM}(a)] confirms the
suppression of neighboring spectral modes within the same cluster.
The suppression of modes in the neighboring clusters was also
tested by measuring the $|g^{(1)}_{s} (\Delta t)|$ for the
filtered signal, see Fig.~\ref{fig:SM}(b). The visibility is above $80\%$
over the $120\, \mathrm{ps}$ range (average $\sim88\pm5\%$) and there is no sign of decay
over this range. This leads to the conclusion that the neighboring
clusters were also suppressed efficiently. The combination of
$G^{(2)}_{s,i} (\tau)$ and $|g^{(1)}_{s} (\Delta t)|$ measurements
is strong evidence of the presence of a single spectral mode in
the signal. Note that only a single filtering stage is needed to
reach the single-mode regime.

The spectral brightness corresponding to the probability of
finding a pair in single-mode fibers (i.e. correcting only for
detection efficiencies) for the filtered source is $11\,
\mathrm{pairs}/\,(\mathrm{s}\cdot\mathrm{mW}\cdot\mathrm{MHz})$. Correcting for all known losses
after the creation of the photon pair we find $8 \times10^3\,
\mathrm{pairs}/\,(\mathrm{s}\cdot\mathrm{mW}\cdot\mathrm{MHz})$, which shows that passive loss is a
major issue to be improved. Compared with the single-pass case,
where we measure a coincidence rate of $3000\,\mathrm{ Hz/mW}$ (for
$\eta_{\mathrm{det,}1436}=0.1$), the cavity enhances the spectral brightness by $B=1250$ \cite{Ou1999} (correcting for filter transmission, cavity escape efficiencies, and chopper duty cycle).

In conclusion, we have demonstrated a quantum light source that is suitable to connect
solid state $\mathrm{Pr}^{3+}$  QMs to fiber optics networks. This is an important step
towards long distance quantum repeaters using solid
state QMs. Beyond this application, our source is also a proof-of-principle that cavity-enhanced down-conversion can be used as a versatile source of narrow-band photon pairs at widely separated
wavelengths, which is relevant in the context of connecting
light-matter quantum interfaces of different kinds.

We would like to thank Florian Wolfgramm, Morgan Mitchell and Majid Ebrahim-Zadeh for valuable discussions.
We are also grateful to IdQuantique and to the European project Q-Essence  for lending us the id220 single-photon counter module. We acknowledge financial support from the European Chistera  QScale Project and ERC Starting Grant QuLIMA.

\appendix
\section{Appendix: Evaluation of the second-order cross-correlation function}
\label{appdx}

The normalized form of $G_{s,i}^{(2)}(\tau)$, the second-order
cross-correlation function between signal and idler fields
($E_{s,i}$), can be expressed as:

\begin{equation}
g_{s,i}^{(2)}(\tau) \equiv \frac{\langle E_s^{\dagger}(t)
E_i^{\dagger}(t+\tau) E_i(t+\tau) E_s(t) \rangle}{\langle
E_i^{\dagger}(t+\tau) E_i(t+\tau) \rangle
\langle E_s^{\dagger}(t) E_s(t) \rangle}, \\
\end{equation}
Following the theory used in \cite{Scholz2009,Wolfgramm2011}, in
the case of doubly-resonant cavity-enhanced downconversion it
takes the form:
\begin{equation}
    \begin{split}
        g_{s,i}^{(2)}(\tau) \propto {} & \Bigg| \sum_{m_s, m_i = 0}^\infty \frac{\sqrt{\gamma_s \, \gamma_i \, \omega_s \, \omega_i}}{\Gamma_s + \Gamma_i} \\
                        & \times
                      \begin{cases}
                                e^{-2 \pi \Gamma_s (\tau-(\tau_0/2))}{\rm sinc}{(i \pi \tau_0 \Gamma_s)} \hspace{3 mm} \hspace{0.5 mm} \tau \geqslant \frac{\tau_0}{2}\\
                                e^{+2 \pi \Gamma_i (\tau-(\tau_0/2))}{\rm sinc}{(i \pi \tau_0 \Gamma_i)} \hspace{4 mm} \hspace{0.5 mm} \tau < \frac{\tau_0}{2}
                      \end{cases} \hspace{-4 mm} \Bigg|^2,
    \end{split}
    \label{eq:G2}
\end{equation}
where $\gamma_{s,i}$ are the cavity damping rates for signal and
idler, $\omega_{s,i}$ are the central frequencies,
 $\Gamma_{s,i} = \gamma_{s,i}/2+i m_{s,i} FSR_{s,i}$ with mode indices $m_{s,i}$ and free spectral ranges $FSR_{s,i}$, and $\tau_0$ is the transit time difference between the signal and idler photons through the SPDC crystal.

\begin{figure}[!htb]
    \centering
    \includegraphics[angle=-90,width=8cm]{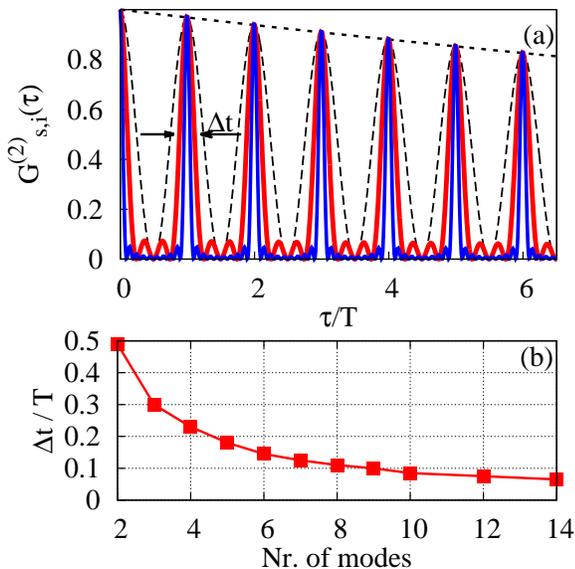}
    \caption{Dependence of the $G^{(2)}_{s,i} (\tau)$ function on the number of spectral modes. (a) Correlation functions for 1, 2, 4, and 10 modes (dotted, dashed, thick solid, and thin solid lines, respectively). (b) Width of the oscillation peaks versus the number of modes.}
     \label{fig:Supp}
\end{figure}

In Fig.~\ref{fig:Supp}(a) simulated $G^{(2)}_{s,i} (\tau)$ functions
are plotted as a function of time delay ($\tau$) normalized to the
cavity roundtrip time ($T$). The different curves correspond to
different number of spectral modes taken into account. 
As the number of modes is increased, the
superposition of the sinc functions results in the oscillation
peak width's ($\Delta t$) reduction. This effect is also shown in
Fig.~\ref{fig:Supp}(b). Taking into account the time resolution of our
detection system ($\sim 685\,\mathrm{ps}$) we can estimate the
expected temporal width of the peaks by convolution. From the best
agreement when changing the number of modes we infer that the
spectrum is composed of clusters containing around 4 longitudinal
modes (peak width $\sim 880\,\mathrm{ps}$). Note that this
corresponds to 4 modes of equal height, which is physically not
the case, but their contribution to the peak width should be
similar.

\bibliographystyle{prsty}

\end{document}